\documentclass[aps,amsmath,amssymb,twocolumn,superscriptaddress]{revtex4}
\usepackage[]{graphicx}
\usepackage{color}
\usepackage{bbold}
\usepackage{braket}
\usepackage[colorlinks=true, pdfstartview=FitV, linkcolor=blue, citecolor=red, urlcolor=black]{hyperref}
\newcommand{\be}{\begin{equation}}
\newcommand{\ee}{\end{equation}}
\newcommand{\ben}{\begin{eqnarray}}
\newcommand{\een}{\end{eqnarray}}
\newcommand{\bes}{\begin{subequations}}
\newcommand{\ees}{\end{subequations}}
\def\bal#1\eal{\begin{align}#1\end{align}}
\usepackage{comment}

\newcommand{\bfi}{\begin{figure}}
\newcommand{\efi}{\end{figure}}
\newcommand{\bc}{\begin{center}}
\newcommand{\ec}{\end{center}}
\newcommand{\sech}{\mbox{sech}}

\begin{document}
\title{Hybrid branes from split kinks}
    \author{D. Bazeia}
    \affiliation{Departamento de F\'\i sica, Universidade Federal da Para\'\i ba, 58051-970 Jo\~ao Pessoa, PB, Brazil}
    \author{A.S. Lob\~ao}
    \affiliation{Escola T\'ecnica de Sa\'ude de Cajazeiras, Universidade Federal de Campina Grande, 58900-000 Cajazeiras, PB, Brazil}    
    \author{M.A. Marques}
    \affiliation{Departamento de Biotecnologia, Universidade Federal da Para\'\i ba, 58051-900 Jo\~ao Pessoa, PB, Brazil}

\begin{abstract}
In this work, we investigate braneworld models generated by scalar fields in which one field has a split kink profile, in which a kink separates into two kinklike configurations. Our analysis covers models with two and three fields, examining the behavior of the most important quantities associated with the brane, such as the warp factor and the stability of the corresponding gravity sector. The results show that the brane is stable and supports a hybrid character, behaving as a thin and thick configuration.
\end{abstract}

\maketitle
\section{Introduction}

Spatially localized structures usually appear as solutions of nonlinear differential equations that play an important role in Physics, as they can be applied in the modeling of specific behavior in high energy physics, in liquid crystals, magnetic materials and other systems \cite{ref1,ref2,ref3,ref4,Vacha}. These objects may also be used in the curved spacetime, within the five-dimensional bulk \cite{ref5,ref6,ref7,ref8,ref9,DeWolfe:1999cp}, where they appear as a source for branes.

In $1+1$ spacetime dimensions, one finds the so-called kinks, which arise under the action of a single scalar field. These structures are static minimum-energy solutions of the equations of motion that connect two neighbor minima of the potential asymptotically. Also, they support a topological charge that protect them against instabilities. Since scalar field models are the simplest ones in Field Theory, they have been vastly used to the search for localized structures exhibiting novel behavior \cite{bnrt1,bnrt2,SV,izquierdo1,izquierdo2,Bazeia:2003qt,alvaro,Afonso:2007mr,Afonso:2008bs,Khare,Bazeia:2008lm}.

In the presence of gravity, scalar fields may be used in the modeling of the fifth dimension in the braneworld scenario. The warp factor associated to the brane proposed in \cite{ref5} engenders a thin profile, with a narrow peak at the origin. The coupling between scalar fields with kinklike profile and gravity was shown to smooth the warp factor of the brane, giving rise to the so-called thick branes which, under some conditions, may be described with first-order equations \cite{ref7,DeWolfe:1999cp}. In this sense, the scalar field solutions impact not only the local geometry of the brane but also its interactions with the surrounding spacetime.

Over the years, several works have appeared in the literature dealing with branes modeled by scalars; see Refs.~\cite{t1,t2,t3,t4,t5,t6,t7,t8,t9,t10,t11,t12,t13,t15,t16}. In particular, in Ref.~\cite{t3}, a two-field model that is capable of describing the degrees of freedom of Bloch walls was used to modeling a thick brane, giving rise to Bloch branes. In Refs.~\cite{t5,t6}, the localization of fermions was investigated, showing that, under specific conditions, chiral fermions could trap the zero modes and resonances related to the internal structure of the brane may appear. In Refs.~\cite{t11,t12}, the authors investigated the effects of the compactification of the scalar field solutions, giving rise to hybrid branes, which behave as thick or thin profile when the extra dimension is inside or outside a compact space. 

Recently, in Ref.~\cite{Bazeia:2024ttl}, a new topological structure called kink crystal was presented in models of two scalar fields. In this structure, the solution is divided into a chain of kinks, where each part individually has a typical topological behavior, connecting minima of the potential. It was observed that this composition of kinks can modify the form of energy density of the system, but does not introduce instabilities in the model, which remains stable under small fluctuations. In this context, one may also find another solution which we call split kink, in which the field configuration appears separated in the form of two kinks, as depicted in Fig. \ref{fig1}.

Motivated by the split kink configuration, this paper aims to investigate how this new structure can modify the behavior of brane models in five spacetime dimensions. To conduct this analysis, we have structured this work as follows: in Section~\ref{sec2}, we provide a detailed description of the mathematical formalism to investigate static configurations in the braneworld scenario. In Section~\ref{sec3} we present two models which are capable of accommodating split kinks. Finally, in Section~\ref{sec4}, we summarize our findings and present perspectives for future investigations.

\section{Methodology}\label{sec2}

Let us start by defining the five-dimensional action that allows the investigation of thick brane models as
\begin{equation}\label{action}
S=\frac1{2\kappa}\int\!\!\sqrt{|g|}\,d^5x\Big(R-2\kappa\,{\cal L}_s\Big),
\end{equation}
where $\kappa$ is a coupling constant, $g$ denotes the determinant of the metric tensor $g_{ab}$, $R$ is the Ricci scalar, and ${\cal L}_s$ represents the Lagrangian density of the scalar fields which act as a source to the brane. In this scenario, Latin indices range from 0 to 4, and Greek indices range from 0 to 3. Moreover, for simplicity, we consider natural units with $4\pi G_5=1$ and set $\kappa=2$. To deal with branes, we consider a Lagrangian density with $N+1$ scalar fields, $\phi$ and $\chi_i$, with $i=1,2,\cdots,N$, in the form
\begin{equation}\label{LagrDens}
{\cal L}_s\!=\!\frac12f(\chi_i)\nabla_{\!a}\phi\nabla^a\phi+\frac12\!\sum_{i=1}^N\nabla_{\!a}\chi_i\nabla^a\chi_i-V(\phi,\chi_i),
\end{equation}
where the field $\phi$ is responsible to introduce the split kink profile and $f(\chi_i)$ is a non-negative function that couples the fields $\chi_i$ with the field $\phi$. By varying the action \eqref{action} with respect to the scalar fields, we obtain
\bes\label{motionphichi}
\bal
&\nabla_a\left(f\nabla^a\phi\right)+V_\phi=0,\label{eqdifphi}\\
&\nabla_a\nabla^a\chi_i-\frac12f_{\chi_i}\nabla_a\phi\nabla^a\phi+V_{\chi_i}=0,\label{eqdifchi}
\eal
\ees
where the indices in $V$ and $f$ represent derivatives with respect to the fields, $V_\phi=\partial V/\partial\phi$, $V_{\chi_i}=\partial V/\partial\chi_i$ and $f_{\chi_i}=df/d\chi_i$. Note that the Eq. \eqref{eqdifchi} is, in fact, a set of $N$ differential equations, one for each field $\chi_i$. Moreover, since the potential in general couples the several fields, the above set of $N+1$ equations are coupled.

By varying the action \eqref{action} with respect to the metric tensor, we obtain the usual Einstein's equation, $G_{ab}=2 T_{ab}$, where $G_{ab}$ represents the Einstein's tensor and $T_{ab}$ is the energy-momentum tensor associated to the source scalar fields. Considering the Lagrangian density \eqref{LagrDens}, we get
\ben\label{TensoEnerMome}
T_{ab}=f\nabla_a\phi\nabla_b\phi+\sum_{i=1}^N\nabla_a\chi_i\nabla_b\chi_i-g_{ab}{\cal L}_s.
\een

The next step in this description is to consider a specific form for the metric tensor $g_{ab}$ that allows for the investigation of brane models. In the context of a five-dimensional braneworld, we take a metric with a flat background, where the warp function $A(y)$ depends solely on the extra dimension $y$, as
\begin{equation}\label{Metric}
    ds^2=e^{2A(y)}\eta_{\mu\nu}dx^\mu dx^\nu-dy^2,
\end{equation}
where $\eta_{\mu\nu}=\text{diag}(+,-,-,-)$ is the Minkowski metric tensor. Additionally, since the warp function depends only on the extra dimension, we take static fields, $\phi=\phi(y)$ and $\chi_i=\chi_i(y)$. Consequently, the equations of motion \eqref{motionphichi} become
\bes\label{EQofMotion}
\bal
&f\phi''+\left(4fA'+\sum_{i=1}^Nf_{\chi_i}\chi_i'\right)\phi'=V_\phi,\\
&\chi_i''+4A'\chi_i'-\frac12\phi'^2f_{\chi_i}=V_{\chi_i},
\eal
\ees
where the prime denotes the derivative with respect to $y$. On the other hand, the non-vanish components of the Einstein equation yield
\bes\label{StaticEinstein}
\bal
A''&=-\frac{2}3\left(f\phi'^2+\sum_{i=1}^N\chi_i'^2\right),\\
A'^2&=\frac{1}{6}\left(f\phi'^2+\sum_{i=1}^N\chi_i'^2\right)-\frac13 V.
\eal
\ees
In principle, for a given $f(\chi_i)$ and $V(\phi,\chi_i)$ we must calculate $\phi(y)$, $\chi_i(y)$ and $A(y)$. Therefore, we have $N+2$ unknown functions that must be obtained from the $N+3$ equations \eqref{EQofMotion} and \eqref{StaticEinstein}. Thus, there is an extra equation which can be shown to be an identity. Obtaining analytical solutions in this problem is in principle a hard task, however, we can adopt the first-order formalism presented in Ref.~\cite{firstbrane} to simplify the problem. We then introduce an auxiliary function $W(\phi,\chi_i)$ such that the potential $V(\phi,\chi_i)$ is expressed as 
\ben\label{FOpotent}
V(\phi,\chi)=\frac12\frac{W_\phi^2}{f}+\frac12\sum_{i=1}^NW_{\chi_i}^2-\frac{4}3W^2.
\een
In this case, the fields and the warp function $A(y)$ are obtained through the first-order equations
\ben\label{FOfields}
\phi'=\frac{W_\phi}{f},\qquad \chi_i'=W_{\chi_i} ,\qquad A'=-\frac{2}3W.
\een
It is important to remark that the first-order equations \eqref{FOpotent} and \eqref{FOfields} are compatible with the equations of motion \eqref{EQofMotion} and Einstein's equations \eqref{StaticEinstein}. The role of the scalar fields as source fields is now clear, as they can be obtained independently from the warp function for a given auxiliary function and then be used to feed $W$ that dictates the form of $A(y)$. The first-order equations allows us to write the energy density in the form
\ben\label{energyDensity}
\rho=\left(We^{2A}\right)'.
\een
This is the necessary formalism to describe the braneworld, so we can now move on to the description of specific models.

\section{Specific Models}\label{sec3}

\begin{figure}[t]
    \begin{center}
    \includegraphics[scale=0.65]{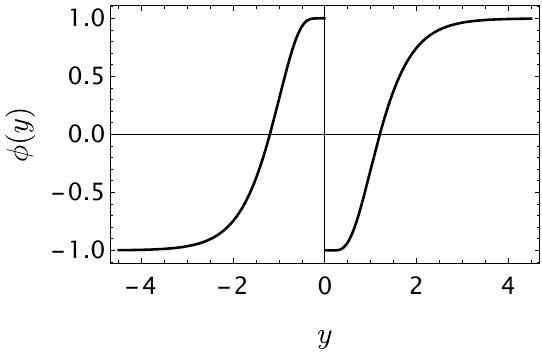}
    \end{center}
    \vspace{-0.5cm}
    \caption{\small{Solution of field $\phi$ for $\alpha=1$.}\label{fig1}}
\end{figure}

We now investigate models that allows for the presence of split kink solutions previously introduced in \cite{Bazeia:2024ttl}. Initially, we consider the case with two scalar fields, $\phi$ and $\chi_1=\chi$, in which $f$ depends only on $\chi$, given by $f(\chi)=\chi^2$, and
\ben\label{W1}
W(\phi,\chi)=\phi-\frac13\phi^3+\alpha\left(\chi-\frac13\chi^3\right).
\een
Here, $\alpha$ represents a real parameter. Although this is a sum of the functions which leads to $\phi^4$ and $\chi^4$ potentials without the presence of couplings between the fields in $W(\phi,\chi)$, the function $f(\chi)$ induces significant effects in the system, modifying the solution of the field $\phi$. In this scenario, the first-order equations for the scalar fields in \eqref{FOfields} take the form
\be\label{fophichi}
\phi'=\frac{1-\phi^2}{\chi^2},\qquad \chi'=\alpha\left(1-\chi^2\right).
\ee
Note that the differential equation for the field $\phi$ is dependent on the field $\chi$. However, the equation for $\chi$ can be solved independently, resulting in $\chi(y)=\tanh(\alpha y+a)$, where $a$ determines the center of the solution. For simplicity, we assume that $a=0$ in this model. Substituting this into the equation for $\phi$, we obtain
\ben\label{solPhi1}
\phi(y)=\tanh\left(y-\frac1\alpha\coth(\alpha y)\right).
\een
The solution for the field $\phi$ exhibits an intriguing behavior, as depicted in Fig.~\ref{fig1}. It engenders a discontinuity at $y=0$ that works to split the kinklike configuration, and behaves asymptotically as a typical exponential tail that appears for the standard kink. Therefore, this structure engender a \emph{split} character. This feature at the origin is only possible due to the presence of the function $f(\chi)$ in the Lagrangian density, which is smooth everywhere. Such split kink solutions were recently proposed in Ref.~\cite{Bazeia:2024ttl}.
Despite the discontinuity of the field $\phi$ at $y=0$, it is of interest to see that the potential $V(\phi,\chi)$ remains finite at the origin, as can be confirmed by
\ben
V(\phi(0),\chi(0))=-\frac{16}{27}+\frac{1}{2}\alpha^2.
\een
Moreover, we have that $V(\phi(\pm\infty),\chi(\pm\infty))=-16(1+\alpha)^2/27$, thus ensuring that the brane is $AdS_5$ .

\begin{figure}[t]
    \begin{center}
    \includegraphics[scale=0.6]{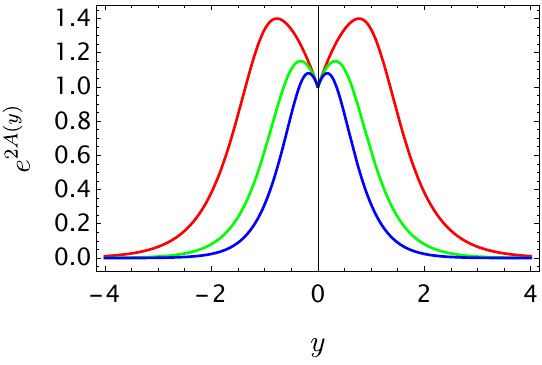}
    \includegraphics[scale=0.6]{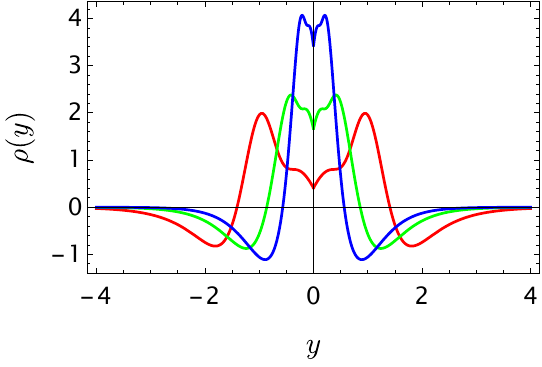}
    \end{center}
    \vspace{-0.5cm}
    \caption{\small{Warp factor (upper panel) and energy density (lower panel), with $\alpha=1.0$ (red line), $\alpha=1.5$ (green line) and $\alpha=2.0$ (blue line).}\label{fig2}}
\end{figure}

An important characteristic of brane is the warp factor, which is defined in terms of the warp function $A(y)$. To obtain this quantity, we can solve the first-order equation for $A'$ provided by the last of the equations in \eqref{FOfields}. It is worth remarking that the behavior of $A(y)$ has a direct relation with the solution of $\phi(y)$ and $\chi(y)$ through the function $W$. Therefore, the discontinuity of the solution at $y=0$ must be dealt with care in the calculation of $A'$. Unfortunately, it's not possible to obtain analytical solutions for the warp function. However, we can use numerical procedures to address the issue. To handle the discontinuity in first derivative of warp function at $y=0$, we split the solution into two regions: the first for the negative and the second for the positive part of $y$. We then perform numerical integration and in the upper panel in Fig.~\ref{fig2} we display the behavior of the warp factor. As anticipated, a discontinuity in the derivative occurs at $y=0$, stemming from the profile of the split kink solution. This result is similar to the thin brane profile, where $A'$ also has discontinuity at $y=0$, see Ref.~\cite{DeWolfe:1999cp}. To better visualize this behavior, we expand the warp factor near to the origin and obtain
\ben
A(y)\approx \frac{4}{9}|y|-\frac{\alpha ^2}{3}y^2.
\een
In essence, the brane exhibits a hybrid behavior, resembling a thin brane in close proximity to the origin and adopting characteristics akin to a thick brane when positioned further away from the origin. Therefore, the split scalar solution \eqref{solPhi1} allows for the interpolation of thin and thick brane profiles in the warp factor. 

\begin{figure}[ht]
    \begin{center}
        \includegraphics[scale=0.6]{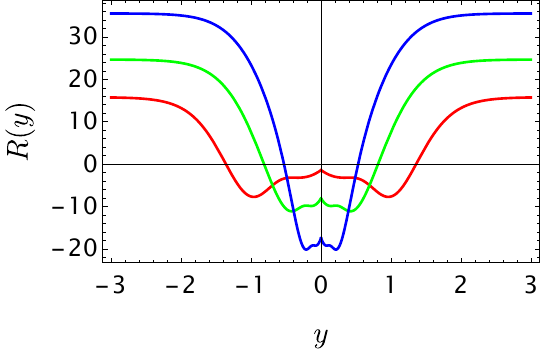}
        \includegraphics[scale=0.6]{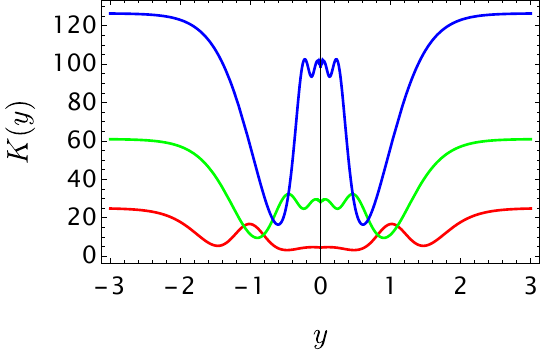}
    \end{center}
    \vspace{-0.5cm}
    \caption{\small{Ricci scalar (upper panel) and Krestschmann scalar (lower panel), with $\alpha=1.0$ (red line), $\alpha=1.5$ (green line) and $\alpha=2.0$ (blue line).}\label{fig3}}
\end{figure}

We also illustrate, in Fig.~\ref{fig2}, the behavior of the energy density for this model, where we also observe the atypical behavior at the origin. In both figures we set the parameter $\alpha=1, 1.5$ and $2$.
In Fig.~\ref{fig3}, we depict the behavior of two quantities strongly influenced by the warp factor, namely the Ricci scalar and the Kretschmann scalar, obtained respectively by the relations,
\bes\label{scalarsCurvatura}
\bal
R&=20A'^2+8A'',\\
K&=40A'^4+32A'^2A''+16A''^2.
\eal
\ees
As expected, these quantities also exhibit a discontinuity in their first derivative at $y=0$, similar to the usual thin brane profile.

We remark that the first-order equation for $\phi$ in \eqref{fophichi} also admits solutions of the half-compact type. These solutions are similar to the one-dimensional kinklike configurations obtained before in Ref.~\cite{geomhc}. For instance, one may consider the solution $\phi(y)=1$ for $y\geq0$ and the expression in Eq.~\eqref{solPhi1} for $y<0$. This solution is continuous, leads to a well-defined brane, but it does not allow for the presence of the internal structure in the warp factor depicted in Fig.~\ref{fig2}.

\begin{figure}[t]
    \begin{center}
    \includegraphics[scale=0.6]{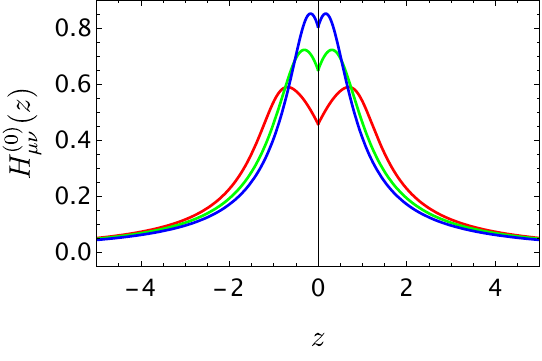}
    \includegraphics[scale=0.6]{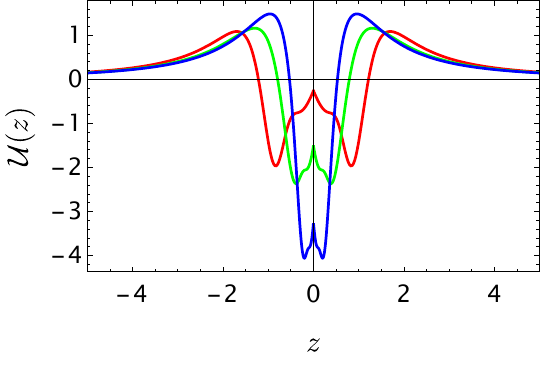}
    \end{center}
    \vspace{-0.5cm}
    \caption{\small{Zero mode (upper panel) and stability potential (lower panel), with $\alpha=1.0$ (red line), $\alpha=1.5$ (green line) and $\alpha=2.0$ (blue line).}\label{fig4}}
\end{figure}

To investigate if the gravity sector of the model defined in Eq.~\eqref{action} is stable against small perturbations in the metric tensor, we follow the lines of Ref. \cite{genbrane}. If we consider $h_{\mu\nu}(x^\sigma,y)$ as a small perturbation in the metric background, we can write that
\ben
ds_5^2=e^{2A}\left(\eta_{\mu\nu}+h_{\mu\nu}\right)dx^\mu dx^\nu-dy^2,
\een
where $h_{\mu\nu}$ obeys the transverse traceless gauge. Introducing the previous metric into Einstein's equation and performing some mathematical manipulations, we can show that the fluctuation obeys a Schrodinger-like equation in the form
\ben
\left(-\frac{d^2}{dz^2}+{\cal U}(z)\right)H_{\mu\nu}(z)=p^2 H_{\mu\nu}(z),
\een
where $z$ represents a new coordinate defined by the relation $dz=e^{-A(y)}dy$. Also, $p$ stands for the eigenvalues of the equation $\Box^{(4)} h_{\mu\nu}=p^2h_{\mu\nu}$, with $H_{\mu\nu}(z)=e^{-ipx}e^{3A(z)/2}h_{\mu\nu}$, and the stability potential ${\cal U}(z)$ is given by
\ben
{\cal U}(z)=\frac94 A_z^2+\frac32 A_{zz}.
\een
Here the index $z$ denotes differentiation with respect to this coordinate. We can also find the zero mode, defined here as $H_{\mu\nu}^{(0)}$, as
\ben
H_{\mu\nu}^{(0)}(z)=N_{\mu\nu} e^{3A(z)/2},
\een
where $N_{\mu\nu}$ is the normalization factor. In the upper panel of Fig.~\ref{fig4}, we depict the zero mode, while in the lower panel, we display the stability potential for the model that gives the solution \eqref{solPhi1}. In both cases, we use $\alpha=1, 1.5$, and $2$. An interesting aspect is that the zero mode displays an internal structure that remains there even when one modifies the values of the $\alpha$ parameter. Moreover, one notices that the zero mode does not engender nodes, ensuring the absence of negative eigenvalues in the stability equation, so the brane is stable.

\begin{figure}[t]
    \begin{center}
    \includegraphics[scale=0.6]{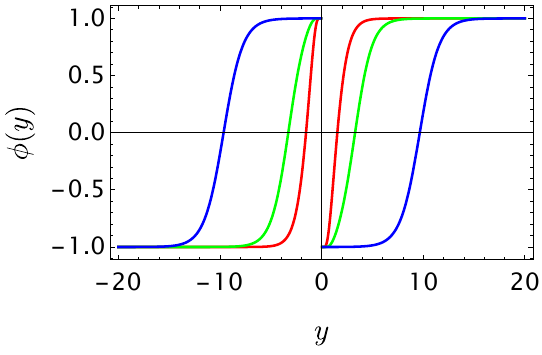}
    \end{center}
    \vspace{-0.5cm}
    \caption{\small{Plots of the split kink solution given by Eq.~\eqref{solPhi2}, with $\alpha=0.6$ (red line), $\alpha=1.2$ (green line) and $\alpha=1.8$ (blue line).}\label{fig5}}
\end{figure}

Next, let us modify the model and introduce a third scalar field, $\chi_2=\psi$, to proceed with the methodology outlined in \cite{t16}. This allows us to investigate the behavior of the internal structure of the brane by shifting the solutions of the fields $\chi$ and $\psi$ with respect to the origin. Let us consider
\ben\label{W2}
W(\phi,\chi,\psi)=\phi\!-\!\frac13\phi^3\!+\!\alpha\!\left(\!\chi\!-\!\frac13\chi^3\!\right)\!+\!\beta\!\left(\!\psi\!-\!\frac13\psi^3\!\right)\!,
\een
where $\beta$ is another real parameter.  Consequently, we can express the solutions for the two fields obtained independently as $\chi(y) = \tanh(\alpha y + a)$ and $\psi(y) = \tanh(\beta y + b)$, where the parameters $a$ and $b$ identify the centers of the two fields $\chi$ and $\psi$. It is of interest to notice that the parameters $\alpha$, $\beta$, $a$, and $b$ may induce an asymmetric behavior in the solutions. This asymmetry in the scalar field profiles may allow for the construction of asymmetric branes \cite{t15,t16}. Here, however, given our intention to assess the impact of the field solutions $\chi(y)$ and $\psi(y)$ on the split kink solution, we opt for taking $\beta=\alpha=1$ and $b=-a$, only considering the symmetric scenario. Furthermore, we consider that the function $f$ depends on both $\chi$ and $\psi$, in the form $f(\chi,\psi)=c+\chi\psi$, where $c$ is a real parameter. Therefore, the first order equation for $\phi$ can be written as follows
\ben
\phi'=\frac{1-\phi^2}{c+\chi\psi}.
\een
To investigate split kink solutions, we take $c=\tanh^2(a)$. By solving this differential equation, we obtain
\ben\label{solPhi2}
\phi(y)=\tanh\big(\xi(y)\big),
\een
where
\ben
\xi(y)\equiv \cosh^2(a) \sech(2a) \left(y-\cosh^2(a)\coth (y)\right).
\een
Fig.~\ref{fig5} illustrates the behavior of the field $\phi$ as the parameter $a$ varies. It is observed that the parameter $a$ modifies the solution near the origin, dispersing or compressing the configuration. Furthermore, if $a=0$, $\phi(y)$ returns to the solution given in Eq.~\eqref{solPhi1} with $\alpha=1$.

\begin{figure}[ht]
    \begin{center}
    \includegraphics[scale=0.6]{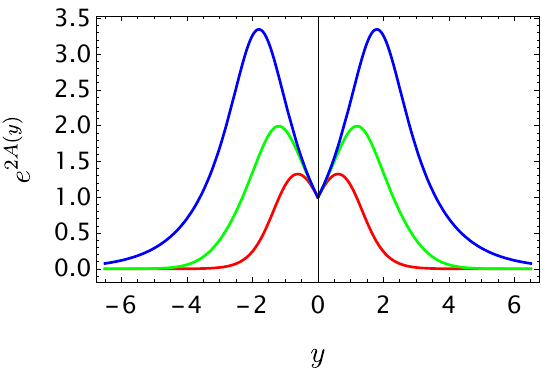}
    \includegraphics[scale=0.6]{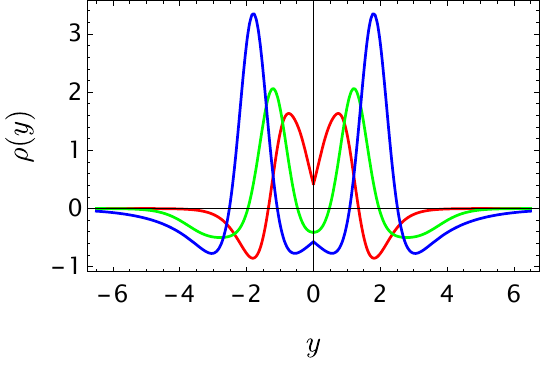}
    \end{center}
    \vspace{-0.5cm}
    \caption{\small{Warp factor (upper panel) and energy density (lower panel), with $\alpha=0.6$ (red line), $\alpha=1.2$ (green line) and $\alpha=1.8$ (blue line).}\label{fig6}}
\end{figure}

\begin{figure}[t]
    \begin{center}
    \includegraphics[scale=0.6]{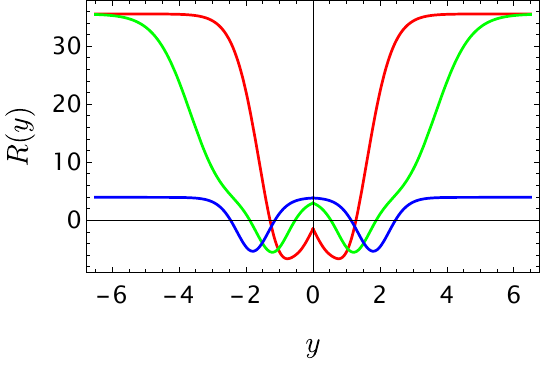}
    \includegraphics[scale=0.6]{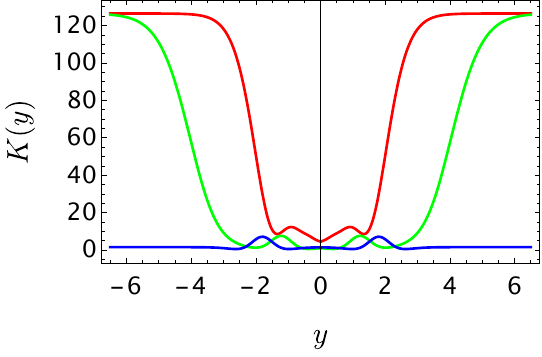}
    \end{center}
    \vspace{-0.5cm}
    \caption{\small{Ricci scalar (upper panel) and Krestschmann scalar (lower panel), with $\alpha=0.6$ (red line), $\alpha=1.2$ (green line) and $\alpha=1.8$ (blue line).}\label{fig7}}
\end{figure}

\begin{figure}[ht]
    \begin{center}
        \includegraphics[scale=0.6]{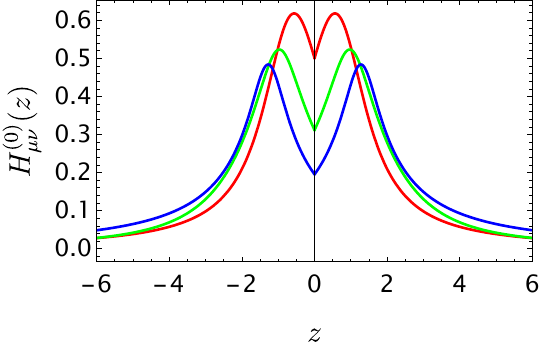}
        \includegraphics[scale=0.6]{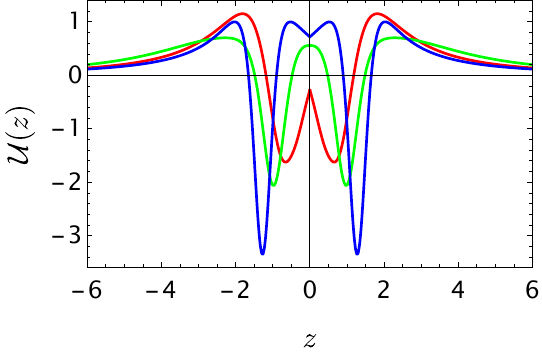}
    \end{center}
    \vspace{-0.5cm}
    \caption{\small{Zero mode (upper panel) and stability potential (lower panel), with $\alpha=0.6$ (red line), $\alpha=1.2$ (green line) and $\alpha=1.8$ (blue line).}\label{fig8}}
\end{figure}

We also analyze the warp factor and energy density in this case, and as in the previous model, only numerical results are possible. These results are represented in Fig.~\ref{fig6}, for some values of the parameter $a$. As observed, the warp factor exhibits a discontinuity in its first derivative at $y=0$.
The Ricci and Krestschmann scalars are presented in Fig.~\ref{fig7}. As in the previous model, both of them have a discontinuity in their first derivative at the origin.  
We also verify the behavior of the zero mode and stability potential in the upper and lower panels of Fig.~\ref{fig8}. We can see that the parameter $a$ changes the maxima and minima outside the origin of these functions. Moreover, one notices that the zero mode has a V-shaped internal structure, which becomes wider and deeper as $a$ increases.

\section{Comments and conclusions}\label{sec4}

In this work we have investigated how the presence of split kinks uncovered in Ref.~\cite{Bazeia:2024ttl} modifies the braneworld scenario that arise in models with two and three scalar fields. In the two-field model, we have shown that the aforementioned solution works to separate the warp factor into two symmetric parts with a minimum and a divergent first derivative at the origin. This behavior also appears in the energy density, Ricci and Kretschmann scalars, as well as in other quantities associated with the stability of the gravitational sector, such as the zero mode and the stability potential. Furthermore, we observe that the zero mode exhibits an internal structure that persists when modifying the values of the parameters.

We have also examined a model comprising three fields, in which two of them engender standard kink solutions with centers symmetrically shifted with respect to the origin. We then have taken these solutions to induce the split kink behavior, whose thickness can be modified by the parameter that controls the separation between the two independent kink solutions. We have observed that this feature also appears in the brane, with the warp factor having a minimum delimited by peaks whose separation is controlled by the the aforementioned parameter. In this three-field model, we have also observed that the split kink does not lead to instabilities in the gravitational sector of the brane. Instead, it results in the emergence of an internal structure in the zero mode, which persists even with the change of parameter values.

As perspectives, one may consider studying branes generated by scalar field models with split kinks in alternative theories of gravity, such as $f(R,T)$ \cite{frt0,frt}, $f(Q)$ \cite{fq}, Teleparallel \cite{tele}, Palatini \cite{palatini} and other scenarios \cite{ho1,ho2,ho3,fr1,fr2,fr3,p1,p2,p3,p4}. We are currently investigating some of these issues and hope to report them in the near future.

\begin{acknowledgments}
The work is supported by the Brazilian agencies Conselho Nacional de Desenvolvimento Cient\'\i fico e Tecnol\'ogico (CNPq), grants No.~303469/2019-6 (DB), No. 402830/2023-7 (DB and MAM) and No. 306151/2022-7 (MAM), and Paraiba State Research Foundation (FAPESQ-PB), grant No.~0015/2019.
\end{acknowledgments}


\end{document}